\documentclass[letter]{article}

\usepackage{amsmath,amsfonts,amssymb,amsgen,amsbsy}
\usepackage{psfig,epsfig,rotating}
\usepackage{listings}

\newcommand{\half}{{\textstyle{1\over2}}}
\newcommand{\third}{{\textstyle{1\over3}}}
\newcommand{\fifth}{{\textstyle{1\over5}}}

\newcommand{\W}{\operatorname{W}}

\title{Efficient resolution of the Colebrook equation}

\author{Didier Clamond}

\date{Laboratoire J.-A. Dieudonn\'e, 06108 Nice cedex 02, France.\\
{\small {\sc E-Mail:} {\sf didierc@unice.fr}}}

\begin{document}
\maketitle

\begin{abstract}
A robust, fast and accurate method for solving the Colebrook-like equations is presented. The algorithm is efficient for the whole range of parameters involved in the Colebrook equation. The computations are not more demanding than simplified approximations, but they are much more accurate. The algorithm is also faster and more robust than the Colebrook solution expressed in term of the Lambert $\W$-function. {\sc Matlab${}^\copyright$} and {\sc FORTRAN} codes are provided.
\end{abstract}

\section{Introduction}\label{secintro}

Turbulent fluid flows in pipes and open channels play an important role in hydraulics, chemical engineering, transportation of hydrocarbons, for example. These flows induce a significant loss of energy depending on the flow regime and the friction on the rigid boundaries. It is thus important to estimate the dissipation due to turbulence and wall friction.

The dissipation models involve a friction coefficient depending on the flow regime (via a Reynolds number) and on the geometry of the pipe or the channel (via an equivalent sand roughness parameter). This friction factor if often given by the well-known Colebrook--White equation, or very similar equations.

The Colebrook--White equation estimates the (dimensionless) Darcy--Weis\-bach
friction factor $\lambda$ for fluid flows in filled pipes. In its most
classical form, the Colebrook--White equation is
\begin{eqnarray}
\frac{1}{\sqrt{\lambda}} &=&-\,2\,\log_{10}\!\left(\,\frac{K}{3.7}\,+\,
\frac{2.51}{R}\frac{1}{\sqrt{\lambda}}\,\right),  \label{col1}
\end{eqnarray}
where $R=UD/\nu$ is a (dimensionless) Reynolds number and $K=\epsilon/D$
is a relative (dimensionless) pipe roughness ($U$ the fluid mean velocity in the pipe, $D$ the pipe hydraulic diameter, $\nu$ the fluid viscosity and $\epsilon$ the pipe absolute roughness height). There exist several variants of the Colebrook equation, e.g.
\begin{eqnarray}
\frac{1}{\sqrt{\lambda}} &=& 1.74\ -\,2\,\log_{10}\!\left(\,2\,K\,+\,
\frac{18.7}{R}\frac{1}{\sqrt{\lambda}}\,\right), \label{col2}\\
\frac{1}{\sqrt{\lambda}} &=& 1.14\ -\,2\,\log_{10}\!\left(\,K\,+\,
\frac{9.3}{R}\frac{1}{\sqrt{\lambda}}\,\right). \label{col3}
\end{eqnarray}
These variants can be recast into the form (\ref{col1}) with small changes in
the numerical constants $2.51$ and $3.7$. Indeed, the latter numbers being obtained fitting experimental data, they are known with limited accuracy. Thus, the formulae (\ref{col2}) and (\ref{col3}) are not fundamentally different from (\ref{col1}). Similarly, there are variants of the Colebrook equations for open channels, which are very similar to (\ref{col1}). Thus, we shall focus on the formula (\ref{col1}), but it is trivial to adapt the resolution procedure introduced here to all variants, as demonstrated in this paper.

The Colebrook equation is transcendent and thus cannot be solved in terms of elementary functions. Some explicit approximate solutions have then been proposed \cite{haa,rom,son2}. For instance, the well-known Haaland formula \cite{haa} reads
\begin{equation}\label{solhaa}
\frac{1}{\sqrt{\lambda}}\ \approx\ -1.81\times\log_{10}\!\left[\,\frac{6.9}{R}\ +\
\left(\frac{K}{3.7}\right)^{1.11}\,\right].
\end{equation}
Haaland's approximation is explicit but is not as simple as it may look. Indeed, this approximation involves one logarithm only, but also a non-integer power. The computation of the latter requires the evaluation of one exponential and one logarithm, since it is generally computed via the relation
$$
x^{1.11}\ =\ \exp(1.11\times\ln(x)),
$$
where `$\ln$' is the natural (Napierian) logarithm. Hence, the overall evaluation of (\ref{solhaa}) requires the computation of three transcendant functions (exponentials and logarithms). We present in this paper much more accurate approximations requiring the evaluation of only two or three logarithms, plus some trivial operations ($+,-,\times,\div$).

Only quite recently, it was noticed that the Colebrook--White equation (\ref{col1}) can be solved in closed form \cite{kea} using the long existing Lambert W-function \cite{cor}. However, when the Reynolds number is large, this exact solution in term of the Lambert function is not convenient for numerical computations due to overflow errors \cite{son}. To overcome this problem, Sonnad and Goudar \cite{son,son2} proposed to combine several approximations depending on the Reynolds number. These approaches are somewhat involve and it is actually possible to develop a simpler and more efficient strategy, as we demonstrate in this paper.

A fast, accurate and robust resolution of the Colebrook equation is, in particular, necessary for scientific intensive computations. For instance, numerical simulations of pipe flows require the computation of the friction coefficient at each grid point and for each time step. For long term simulations of long pipes, the Colebrook equation must therefore be solved a huge number of times and hence a fast algorithm is required. An example of such demanding code is the program {\sc OLGA} \cite{olga} which is widely used in the oil industry.

Although the Colebrook formula itself is not very accurate, its accurate resolution is nonetheless an issue for numerical simulations because a too crude resolution may affect the repeatability of the simulations. Robustness is also important since one understandably wants an algorithm capable of dealing with all the possible values of the physical parameters involved in the model. The method described in the present paper was developed to address all this issues. It is also very simple so it can be used for simple applications as well. The method proposed here aims at giving a definitive answer to the problem of solving numerically the Colebrook-like equations.

The paper is organized as follow. In section \ref{secgencol}, a general Colebrook-like equation and its solution in term of the Lambert $\W$-function are presented. For the sake of completeness, the Lambert function is briefly described in section \ref{secWfun}, as well as a standard algorithm used for its computation. A severe drawback of using the Lambert function for solving the Colebrook equation is also pointed out. To overcome this problem, a new function is introduced in section \ref{secOfun} and an improved new numerical procedure is described. Though this function introduces a big improvement for the computation of the friction factor, it is still not fully satisfactory for solving the Colebrook equation. The reasons are explained in the section \ref{secpifun}, where a modified function is derived to address the issue. The modified function is subsequently used in section \ref{secsolcol} to solve the Colebrook equation efficiently. The accuracy and speed of the new algorithm is tested and compared with Haaland's approximation. For testing the method and for intensive practical applications, {\sc Matlab${}^\copyright$} and {\sc FORTRAN} implementations of the algorithm are provided in the appendices. The algorithm is so simple that it can easily be implemented in any other language and modified to be adapted to any variants of the Colebrook equation.

\section{Generic Colebrook equation and its solution}\label{secgencol}

We consider here a generic Colebrook-like equation as
\begin{equation}\label{colgen}
\frac{1}{\sqrt{\lambda}}\ =\ c_0\ -\ c_1\,\ln\!\left(\,c_2\,+\,
\frac{c_3}{\sqrt{\lambda}}\,\right),
\end{equation}
where the $c_i$ are given constants such that $c_1c_3>0$. The classical Colebrook--White formula (\ref{col1}) is obviously obtained as a special case of (\ref{colgen}) with $c_0=0$, $c_1=2/\ln10$, $c_2=K/3.7$ and $c_3=2.51/R$.

The equation (\ref{colgen}) has the exact analytical solution
\begin{equation}\label{colgensolW}
\frac{1}{\sqrt{\lambda}}\ =\ c_1\left[\,\W\!\left(\exp\left(\,{c_0\over
c_1}+{c_2\over c_1\/c_3} - \ln(c_1\/c_3)\right)\,\right)\ -\ \frac{c_2}{c_1\/c_3}\,\right],
\end{equation}
which is real if $c_1c_3>0$ and where $\W$ is the principal branch of the Lambert function, often denoted $\W_0$ \cite{cor}. In this paper, only the principal branch of the Lambert function is considered because the other branches correspond to non-physical solutions of the Colebrook equations, so the simplified notation $\W$ is not ambiguous.

\section{Brief introduction to the Lambert $\W$-function}\label{secWfun}

For the sake of completeness, we briefly introduce the Lambert function and its practical computation. Much more details can be found in \cite{cor,ser}.

The Lambert $\W$-function solves the equation
\begin{eqnarray} \label{eqW}
y\,\exp(y)\ =\ x \qquad \Longrightarrow \qquad y\ =\ \W(x),
\end{eqnarray}
where, here, $x$ is real --- more precisely $x\geqslant-\exp(-1)$ --- and $\W(0)=0$. The Lambert function cannot be expressed in terms of elementary functions. An efficient algorithm for its computation is based on Halley's iterations \cite{cor}
\begin{equation} \label{halori}
y_{j+1}\ =\ y_j\ -\ \frac{y_j\,\exp(y_j)\,-\,x} {(y_j+1)\,\exp(y_j)\,-\,
\half(y_j+2)\,(y_j\,\exp(y_j)-x)/(y_j+1)},
\end{equation}
provided an initial guess $y_0$. Halley's method is cubic (c.f. Appendix \ref{appsoleq}), meaning that the number of exact digits is (roughly) multiplied by three after each iteration. Today, programs for computing the Lambert function are easily found. For instance, an efficient implementation in {\sc Matlab${}^\copyright$} (including complex argument and all the branches) is freely available \cite{get}.

The Taylor expansion around $\,x=0\,$ of the Lambert function  is
\begin{equation}\label{expW0}
\mathrm{W}(x)\ =\ \sum_{n=1}^\infty\,\frac{(-n)^{n-1}}{n!}\,x^n, \qquad
|x|<\exp(-1).
\end{equation}
This expansion is of little interest to solve the Colebrook equation because, in this context, the corresponding variable $x$ is necessarily large ($x\gg1$). It is thus more relevant to consider the asymptotic expansion
\begin{eqnarray}\label{devWexp}
\mathrm{W}(x)\ \sim\ \ln(x)\ -\ \ln(\ln(x)) \qquad
\text{as}\quad x\rightarrow\infty.
\end{eqnarray}
This expansion reveals that $\W$ behaves logarithmically for large $x$, while we must compute $\W(\exp(x))$ to solve the Colebrook equation, c.f. relation (\ref{colgensolW}). For our applications, $x$ is large and $\exp(x)$ is therefore necessarily huge, to an extend that the computation of $\exp(x)$ cannot be achieved due to overflow. Even when the intermediate computations can be done, the result can be very inaccurate due to large round-off errors. Therefore, the resolution of the Colebrook equation via the Lambert function \cite{kea} is not efficient for the whole range of parameter of practical interest \cite{son}.

\section{The $\omega$-function}\label{secOfun}

To overcome the numerical difficulties related to the Lambert $\W$-function, when used for solving the Colebrook--White equation, we introduce here a new function: the $\omega$-function.

The $\omega$-function is defined such that it solves the equation
\begin{eqnarray} \label{eqom}
y\ +\ \ln(y)\ =\ x \qquad \Longrightarrow \qquad y\ =\ \omega(x),
\end{eqnarray}
where we consider only real $x$. The $\omega$-function is related to the $\W$-function as
\[
\omega(x)\ =\ \W(\exp(x)).
\]
Note that the Lambert $\W$-function is also sometimes called the Omega function, that should not be confused with the $\omega$-function defined here, where we follow the notation used in \cite{orc}. In terms of the $\omega$-function, the solution of (\ref{colgen}) is of course
\begin{equation}\label{colgensolome}
\frac{1}{\sqrt{\lambda}}\ =\ c_1\left[\,\omega\!\left(\,{c_0\over
c_1}+{c_2\over c_1\/c_3}-\ln(c_1\/c_3)\right)\ -\ \frac{c_2}{c_1\/c_3}\,\right].
\end{equation}
For large arguments $\omega(x)$ behaves like $x$, i.e. we have the asymptotic behavior
\begin{eqnarray}\label{devOexp}
\omega(x)\ \sim\ x\ -\ \ln(x) \qquad
\text{as}\quad x\rightarrow\infty,
\end{eqnarray}
which is an interesting feature for the application considered in this paper.

\begin{figure}
\vspace{5cm} \makebox{ \setlength{\unitlength}{1mm}
\begin{picture}(140,90)(0,0)
\put(0,0){\psfig{figure=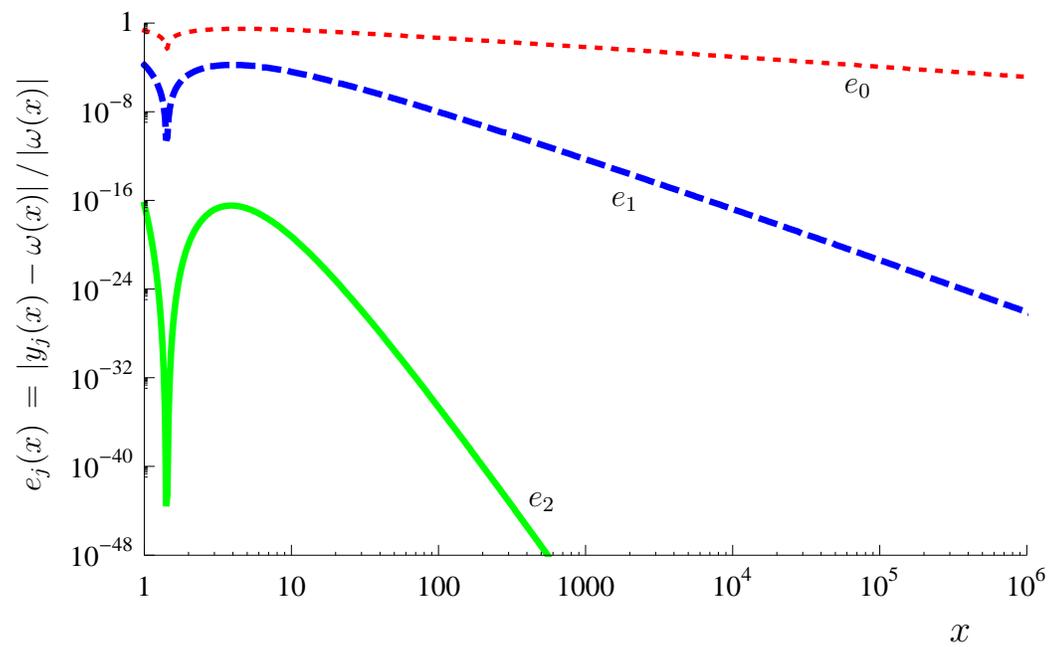,height=8cm,width=13cm}}
\put(117,-5){\Large $x$}
\put(61,13){\large $e_2$}\put(72,53){\large $e_1$}\put(103,68){\large $e_0$}
\put(-7,15){\rotatebox{90}{\large $e_j(x)\,=\,|y_j(x)-\omega(x)|\,/\,|\omega(x)|$}}
\end{picture}}
\vspace{1cm} \caption{\label{figerrome}{\em Relative errors $e_j$ of the $\omega$-function computed via the iterations (\ref{solwexp}).}}
\begin{center}
{Dotted red line: $e_0$; Dashed blue line: $e_1$; Solid green line: $e_2$.}
\end{center}
\end{figure}

As noted by Corless {\em et al.} \cite{ser}, the equation (\ref{eqom}) is in some ways nicer than (\ref{eqW}). In particular, its derivatives (with respect of $y$) are simpler, leading thus to algebraically simpler formulae for its numerical resolution. An efficient iterative quartic scheme (c.f. Appendix \ref{appsoleq}) is thus
\begin{equation}\label{solwexp}
y_{j+1}\ =\ y_j \ -\
\frac{\left(\,1+y_j+\half\/\epsilon_j\,\right)\epsilon_j\,y_j}
{\left(\,1+y_j+\epsilon_j+\third\/\epsilon_j^{\,2}\,\right)}
\qquad \mathrm{for}\quad j\geqslant1,
\end{equation}
with
\[
\epsilon_j\ \equiv\ \frac{y_j\,+\,\ln(y_j)\,-\,x}{1\,+\,y_j}, \qquad
y_0\ =\ x\ -\ \frac{1}{5}.
\]
The computationally costless initial guess ($y_0=x-\fifth$) was obtained considering the asymptotic behavior (\ref{devWexp}), minus an empirically found correction (the term $-\fifth$) to improve somewhat the accuracy of $y_0$ for small $x$ without affecting the accuracy for large $x$.
The relative error $e_j$ of the $j$-th iteration, i.e.
\[
e_j(x)\ \equiv\ \left|\,\frac{y_j(x)\,-\,\omega(x)}{\omega(x)}\,\right|,
\]
is displayed on the figure \ref{figerrome} for $1\leqslant x\leqslant10^6$ and $j=0,1,2$. (The accuracy of (\ref{solwexp}) were measured using the arbitrary precision capability of {\sc Mathematica${}^\copyright$}.) We can see that with $j=2$ we have already reached the maximum accuracy possible when computing in double precision, since $\max(e_2)\approx4\times10^{-17}$ for $x\in[1;\infty[$. We note that the relative error continues to decay monotonically as $x$ increases (even for $x>10^6$) and that there are no overflow problems when computing $y_j$ even for very large $x$ (i.e. $x\gg10^6$). We note also that for $x\gtrapprox5700$ the machine double precision is obtained after one iteration only.

The scheme (\ref{solwexp}) is quartic, meaning that the number of exact digits is multiplied by four after each iteration (c.f. Appendix \ref{appsoleq}). Hence, starting with an initial guess with one correct digit, four digits are exact after one iteration and sixteen after two iterations. That is to say that the machine precision (if working in double precision) is achieved after two iterations only (Fig. \ref{figerrome}). Moreover, the scheme (\ref{solwexp}) has a comparable algebraic complexity per iteration than the scheme (\ref{halori}), i.e. the computational times per iteration are almost identical. However, the iterative quartic scheme (\ref{solwexp}) converges faster than the cubic one (\ref{halori}), and there are no overflow problems as they appear when computing $\W(\exp(x))$ for large $x$. This algorithm could therefore be used to compute the solution of the Colebrook--White equation (\ref{col1}), but we will use instead an even better one defined in the next section. We note in passing that the iterations (\ref{solwexp}) are also efficient for computing the $\omega$-function for any complex $x$, provided some changes in the initial guess $y_0$ depending on $x$.\\

{\bf Remarks:}

{\em i}- With a more accurate initial guess $y_0$, such as $y_0=x-\ln(x)$, the desired accuracy may be obtained with fewer iterations. However, the computation of such an improved initial guess requires the evaluation of transcendent functions. Thus, it cannot be significantly faster than the evaluation of $y_1$ with (\ref{solwexp}) from the simplest guess $y_0=x-\fifth$, and most likely less accurate.

{\em ii}- Higher-order iterations are generally more involved per iteration than the low-order ones. Higher-order iterations are thus interesting if the number of iterations is sufficiently reduced so that the total computation is faster to achieve the desired accuracy. This is precisely the case here.

{\em iii}- Intensive tests have convinced us that the choice of the simplest initial guess $y_0=x-\fifth$ together with the quartic iterations (\ref{solwexp}) is probably the best possible scheme for computing the $\omega$-function in the interval $x\in[1;\infty[$, at least when working in double precision. If improvements can be found, they are thus most likely very minor in terms of both robustness, speed and accuracy.

\section{The $\varpi$-function}\label{secpifun}

Solving the Colebrook equation via the $\omega$-function is a big improvement compared to its solution in term of the Lambert $\W$-function. One can check
that the numerical resolution of the Colebrook equation via the algorithm (\ref{solwexp}) is indeed very efficient when $K=0$, even for very large $R$. However, when $K>0$ the scheme (\ref{solwexp}) is not so effective for large $R$, meaning that not all the numerical shortcomings have been addressed introducing the $\omega$-function. The cause for these numerical problems can be explained as follow.

The solution of the Colebrook equation requires the computation of an expression like $\omega(x_1+x_2)-x_1$ where $x_1\gg x_2$  when $R$ is large and $K\neq0$ (but $x_1=0$ if $K=0$), see the relation (\ref{colsolome}) below. Assuming $x_2\propto\ln(x_1)$, as is the case here, the asymptotic expansion as $x_1\rightarrow\infty$, i.e.
\[
\omega(x_1+x_2)\,-\,x_1\ \sim\ (x_1+x_2-\ln(x_1))\,-\,x_1\ =\ x_2\,-\,\ln(x_1),
\]
exhibits the source of the numerical problems. Indeed, when $K>0$ and $R$ is large, we have $x_1\gg x_2$ and $x_1\gg\ln(x_1)$. Therefore $|x_2-\ln(x_1)|/x_1$ can be smaller than the accuracy used in the computation and we thus obtain numerically $x_1+x_2-\ln(x_1)\approx x_1$ due to round-off errors. Hence $\omega(x_1+x_2)-x_1\approx0$  is computed instead of $\omega(x_1+x_2)-x_1\approx x_2-\ln(x_1)$. To overcome this problem we introduce yet another function: the $\varpi$-function.

Introducing the change of variable $y=z+x_1$ into the equation (\ref{eqom}), the $\varpi$-function is defined such that it solves the equation
\begin{eqnarray} \label{eqpi}
z\ +\ \ln(x_1+z)\ =\ x_2 \qquad \Longrightarrow \qquad z\ =\ \varpi(x_1\left|\,x_2\right.),
\end{eqnarray}
where the $x_i$ are real. The $\varpi$-function is related to the $\omega$- and $\W$-functions as
\[
\varpi(x_1\left|\,x_2\right.)\ =\ \omega(x_1+x_2)\,-\,x_1\ =\ \W(\exp(x_1+x_2))\,-\,x_1.
\]
In terms of the $\varpi$-function, the solution of (\ref{colgen}) is obviously
\begin{equation}\label{colgensolvpi}
\frac{1}{\sqrt{\lambda}}\ =\ c_1\,\varpi\!\left(\,{c_2\over c_1\/c_3}\left|\,{c_0\over c_1}-\ln(c_1\/c_3)\right.\right).
\end{equation}
The  $\varpi$-function is nothing more than the  $\omega$-function shifted by the quantity $x_1$. This is a very minor analytic modification but this is a numerical significant improvement when $x_1$ is large.

An efficient numerical algorithm for computing the $\varpi$-function is directly derived from the scheme (\ref{solwexp}) used for the $\omega$-function. We thus obtain at once
\begin{equation}\label{solvpi}
z_{j+1}\ =\ z_j \ -\
\frac{\left(\,1+x_1+z_j+\half\/\epsilon_j\,\right)\epsilon_j\,(x_1+z_j)}
{\left(\,1+x_1+z_j+\epsilon_j+\third\/\epsilon_j^{\,2}\,\right)}
\qquad \mathrm{for}\quad j\geqslant1,
\end{equation}
with
\[
\epsilon_j\ \equiv\ \frac{z_j\,+\,\ln(x_1+z_j)\,-\,x_2}{1\,+\,x_1\,+\,z_j}, \qquad
z_0\ =\ x_2\ -\ \frac{1}{5}.
\]
If $x_1=0$ the scheme (\ref{solwexp}) is recovered. The rate of convergence of (\ref{solvpi}) is of course identical to the scheme (\ref{solwexp}). Thus, the efficiency of (\ref{solvpi}) does not need to be re-discussed here (see section \ref{secOfun}).

\section{Resolution of the Colebrook--White equation}\label{secsolcol}

We test the new procedure with the peculiar Colebrook--White equation (\ref{col1}). Its general solution is
\begin{eqnarray}
\frac{1}{\sqrt{\lambda}} &=& \frac{2}{\ell}\left[\,\W\left(\exp\left(\frac{\ell\,K\, R}{18.574}\,+\,\ln\!\left(\frac{\ell\,R}{5.02}\right)\right)\right)\ -\ \frac{\ell\,K\, R}{18.574}\,\right]\label{colsollam} \\
&=&\frac{2}{\ell}\left[\,\operatorname{\omega}\!\left(\frac{\ell\,K\, R}{18.574}\,+\,\ln\!\left(\frac{\ell\,R}{5.02}\right)\right)\ -\ \frac{\ell\,K\, R}{18.574}\,\right]\label{colsolome}\\
&=&\frac{2}{\ell}\,\operatorname{\varpi}\!\left(\frac{\ell\,K\, R}{18.574}\,\left|\,\ln\!\left(\frac{\ell\,R}{5.02}\right)\right.\right),\label{colsolvpi}
\end{eqnarray}
where $\ell=\ln(10)\approx2.302585093$. All these analytic solutions are mathematically equivalent, but the relation (\ref{colsolvpi}) is more efficient for numerical computations if we use the scheme described in the previous section.

\subsection{Numerical procedure}

The solution of the Colebrook--White equation is obtained computing the  $\varpi$-function with
\[
x_1\ =\ \frac{\ell\,K\,R}{18.574}, \qquad
x_2\ =\ \ln\!\left(\,\frac{\ell\,R}{5.02}\,\right),
\]
and using the iterative scheme (\ref{solvpi}) with $j=0,1,2$. An approximation of the friction factor is eventually
\[
\lambda_j\ \approx\ (\,\ell\,/\,2\,z_j\,)^2.
\]
This way, the whole computation of $\lambda_j$ requires the evaluation of $j+1$ logarithms only,\footnote{The numerical constant $\ln(10)$ is not counted because it can be explicitly given in the program and does not need to be computed each time.} i.e. one logarithm per iteration.

A {\sc Matlab${}^\copyright$} implementation of this algorithm is given in the appendix \ref{appmat}. This (vectorized) code was written with clarity in mind, so that one can test and modify easily the program. This program is also fast, accurate and robust, so it can be used in real intensive applications developed in {\sc Matlab}.

A {\sc FORTRAN} implementation of this algorithm is given in the appendix \ref{appfor}. This program was written with speed in mind, so there are no checks of the input parameters. The code is clear enough that it should be easy to modify and to translate into any programming language.

\subsection{Accuracy}

\begin{figure}
\makebox{ \setlength{\unitlength}{1mm}
\begin{picture}(160,110)(0,0)
\put(00,55){\psfig{figure=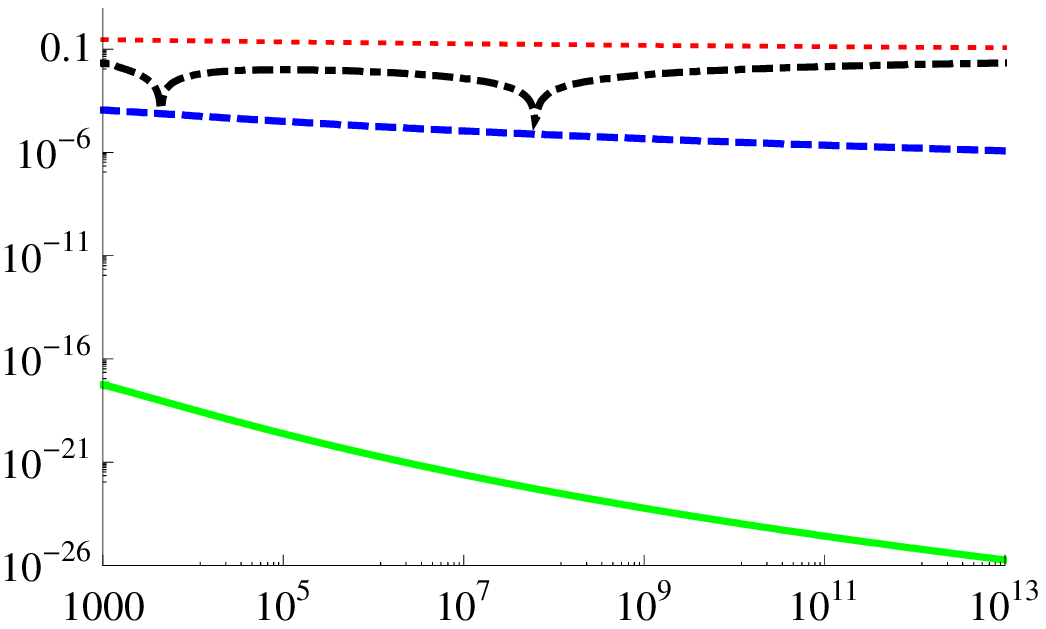,height=5cm,width=6cm}}
\put(65,55){\psfig{figure=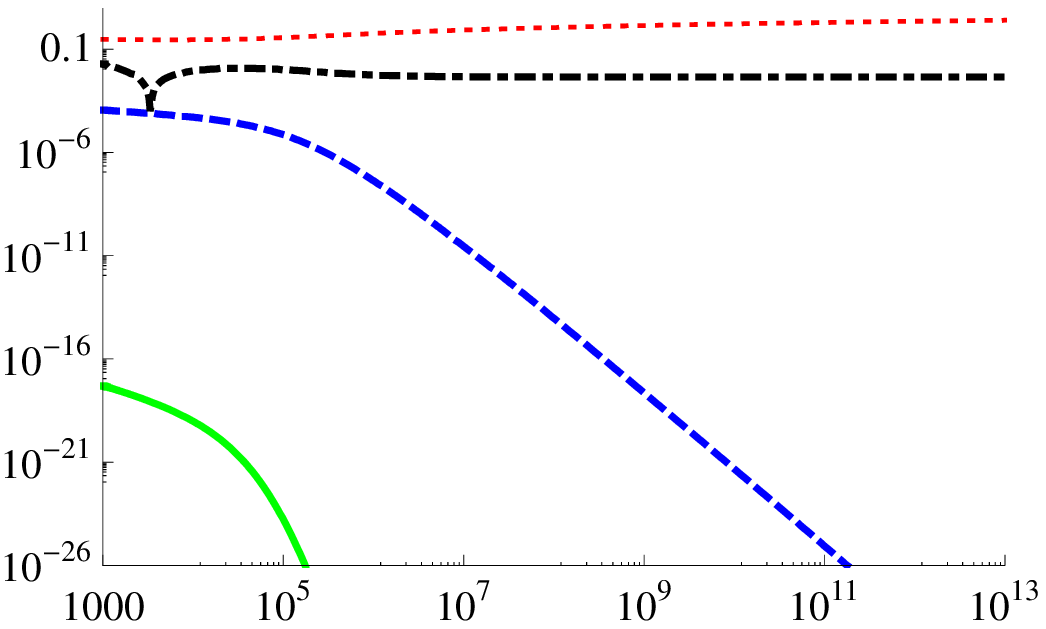,height=5cm,width=6cm}}
\put(00,0){\psfig{figure=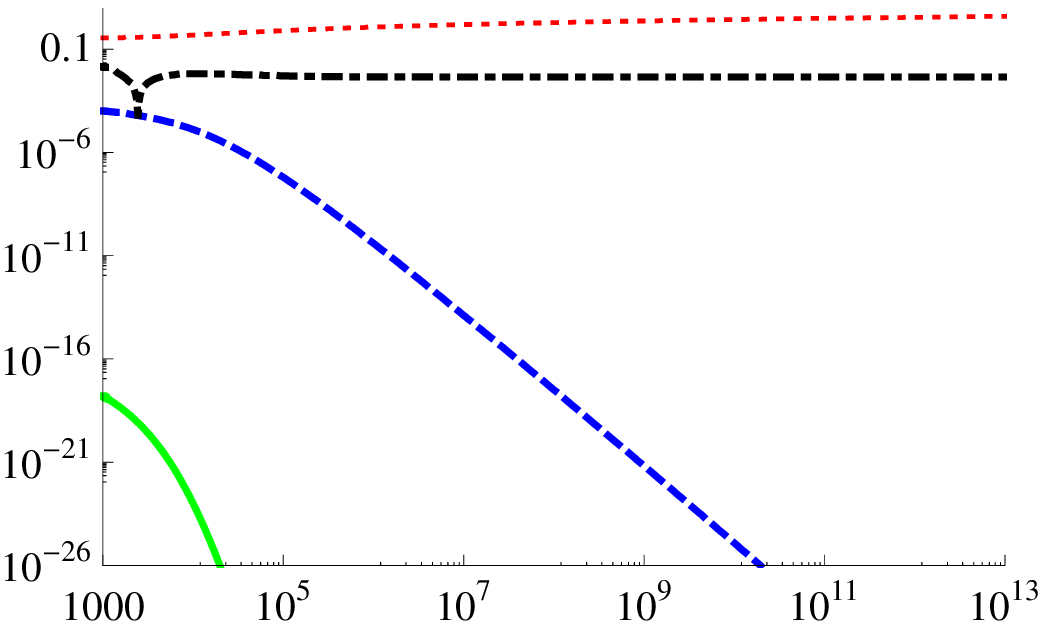,height=5cm,width=6cm}}
\put(65,00){\psfig{figure=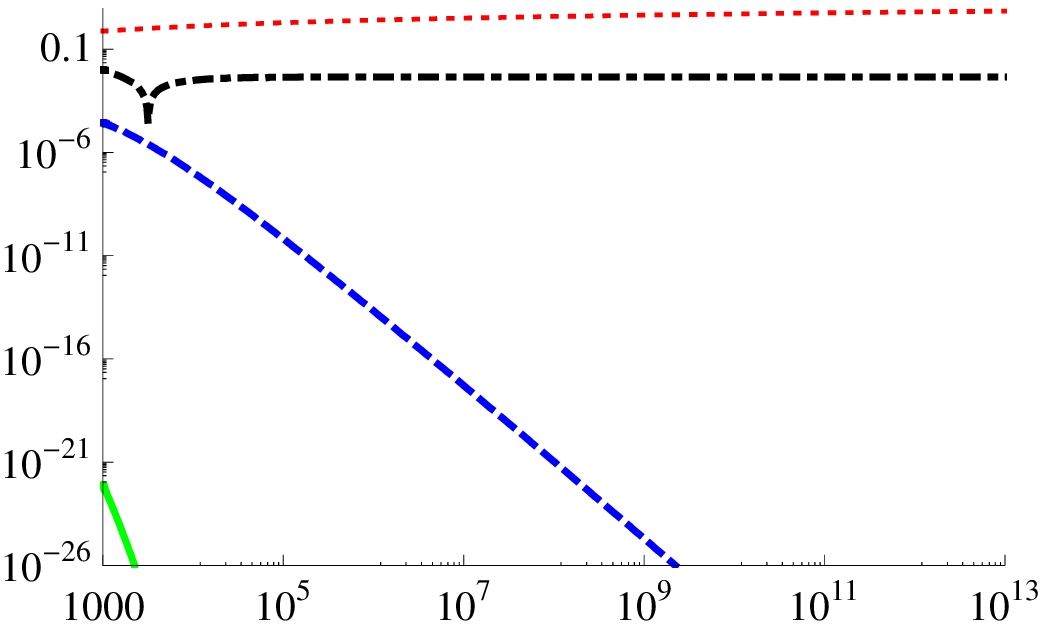,height=5cm,width=6cm}}
\put(95,-5){\large $R$}
\put(30,-5){\large $R$}
\put(-7,65){\rotatebox{90}{$|\lambda_j^{-{1\over2}}-\lambda^{-{1\over2}}|\,/\,|\lambda^{-{1\over2}}|$}}
\put(-7,10){\rotatebox{90}{$|\lambda_j^{-{1\over2}}-\lambda^{-{1\over2}}|\,/\,|\lambda^{-{1\over2}}|$}}
\end{picture}}
\vspace{1cm} \caption{\label{figerrcole}{\em Relative errors of $\lambda^{-{1\over2}}$, computed via the iterations (\ref{solvpi}) and the Haaland formula (\ref{solhaa}), as functions of the Reynolds number $R$. Upper-left: $K=0$; Upper-right: $K=10^{-3}$; Lower-left: $K=10^{-2}$; Lower-right: $K=10^{-1}$.}}
\begin{center}
{Dotted red line: $\lambda_0^{-{1\over2}}$; Dashed blue line: $\lambda_1^{-{1\over2}}$; Solid green line: $\lambda_2^{-{1\over2}}$; Dashed-dotted black line $\lambda_\text{H}^{-{1\over2}}$ (Haaland's approximation).}
\end{center}
\end{figure}

For the range of Reynolds numbers $10^3\leqslant R\leqslant 10^{13}$ and for four relative roughness $K=\{0,10^{-3},10^{-2},10^{-1}\}$, the accuracy of $\lambda_j^{-1/2}$ --- obtained from the iterations (\ref{solvpi}) with $j=\{0,1,2\}$ --- and of Haaland's approximation $\lambda_\text{H}^{-1/2}$ --- given by (\ref{solhaa}) --- are compared with the exact friction coefficient $\lambda^{-1/2}$. The relative errors are displayed on the figure \ref{figerrcole}.

It appears clearly that $\lambda_2^{-1/2}$ is accurate to machine double-precision (at least) for all Reynolds numbers and for all roughnesses (in the whole range of physical interest, and beyond).

It also appears that  $\lambda_1^{-1/2}$ is more accurate than Haaland's approximation, specially for large $R$ and $K$. Moreover, the computation of $\lambda_1$ requires the evaluation of only two logarithms, so it is faster than Haaland's formula.  Note that other explicit approximations having more or less the same accuracy as Haaland's formula, $\lambda_1^{-1/2}$ is significantly more accurate than these approximations.

Finally, we note that $\lambda_0^{-1/2}$ is a too poor approximation to be of any practical interest.

\subsection{Speed}

Testing the actual speed of an algorithm is a delicate task because the running time depends of many factors independent of the algorithm itself (implementation, system, compiler, hardware, etc.), specially on multi-tasking and multi-users computers. In order to estimate the speed of our scheme as fairly as possible, the following methodology was used.

The speeds of the computation of $\lambda_1$ and $\lambda_2$ are compared with the Haaland approximation $\lambda_\text{H}$. The {\sc Matlab} environnement and its built-in {\sf cputime} function is used, for simplicity.

\begin{figure}
\makebox{ \setlength{\unitlength}{1mm}
\begin{picture}(120,80)(0,0)
\put(-5,00){\psfig{figure=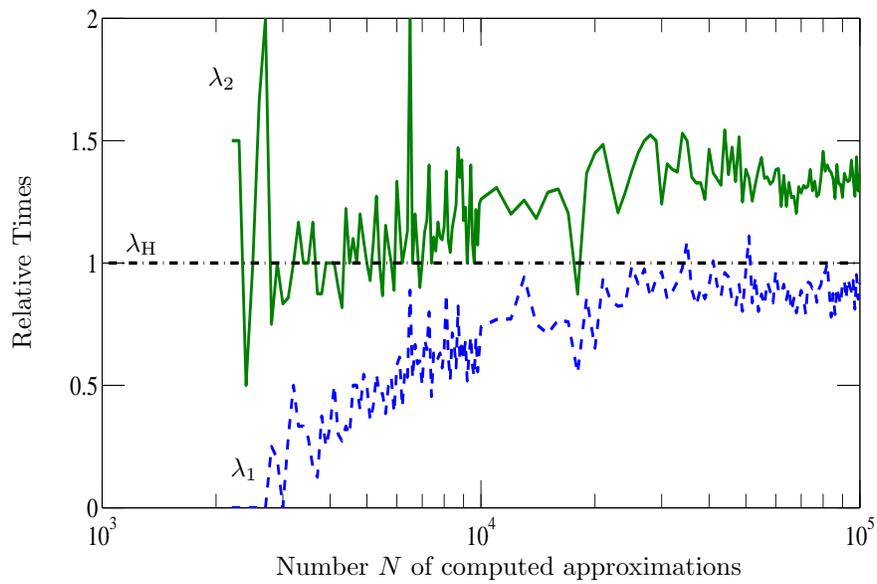,height=8cm,width=13cm}}
\put(35,-0){\normalsize Number $N$ of computed approximations}
\put(-0,30){\rotatebox{90}{Relative Times}}
\put(15,43){$\lambda_{{\tiny\text{H}}}$}
\put(29,13){$\lambda_1$}
\put(26,65){$\lambda_2$}
\end{picture}}
\vspace{1cm} \caption{\label{figspecole}{\em Computational times of $\lambda_1$ (dashed blue line) and $\lambda_2$ (solid green line) with respect of the Haaland approximation $\lambda_\text{H}$ (dashed-dotted black line).}}
\end{figure}

Two vectors of $N$ components, with $1\leqslant N\leqslant10^5$, are created for $R$ and $K$. The values are chosen randomly in the intervals $10^3\leqslant R\leqslant10^9$ and $0\leqslant K<1$. The computational times are measured several times, the different procedures being called in different orders. For each value of $N$, the respective timings are averaged and divided by the averaged time used by the Haaland approximation (the latter having thus a relative computational time equal to one for all $N$). The result of this test are displayed on the figure \ref{figspecole}. (The whole procedure was repeated several times and the corresponding graphics were similar.)

For small $N$, say $N<2000$, the computations are so fast that the function {\sf cputime} cannot measure the times. For larger values of $N$, we can see on the figure \ref{figspecole} that the computations of $\lambda_1$ are a bit faster than the Haaland formula, while the computations of $\lambda_2$ are a bit
slower, in average. This is in agreement with the number of evaluations of transcendent functions needed for each approximations, as mentioned above.

These relative times may vary depending on the system, hardware and software, but we believe that the results would not be fundamentally different from the ones obtained here. The important result is that the procedure presented in this paper is comparable, in term of speed, to simplified formulae such as the Haaland approximation. The new procedure being much more accurate, it should thus be preferred.

\section{Conclusion}

We have introduced a simple, fast, accurate and robust algorithm for solving the Colebrook equation. The formula used is the same for the whole range of the parameters. The accuracy is around machine double precision (around sixteen digits). The present algorithm is more efficient than the solution of the Colebrook equation expressed in term of the Lambert $\W$-function and than simple approximations, such as the Haaland formula.

We have also provided routines in {\sc Matlab} and {\sc FORTRAN} for its practical use. The algorithm is so simple that it can easily be implemented in any other language and can be adapted to any variant of the Colebrook equation.

To derive the algorithm, we introduced two special functions: the $\omega$- and $\varpi$-functions. These functions could also be useful in other contexts than the Colebrook equation. The efficient algorithms introduced in this paper for their numerical computation could then be used, perhaps with some modifications of the initial guesses, specially if high accuracy is needed.

\appendix

\section{High-order schemes for solving a single nonlinear equation}\label{appsoleq}

Let be a single nonlinear equation $f(y)=0$, where $f$ is a sufficiently regular given function and $y$ is unknown. This equation can be solved iteratively via the numerical scheme \cite{hou}
\begin{equation}\label{defhou}
y_{j+1}\ =\ y_j\ +\
(p+1)\left[\frac{(1/f)^{(p)}}{(1/f)^{(p+1)}}\right]_{y=y_j},
\end{equation}
where $p$ is a non-negative integer and $F^{(p)}$ denotes the $p$-th derivative of $F$ with $F^{(0)}=F$.

The scheme (\ref{defhou}) is of order $p+2$, meaning that the number of exact digits is roughly multiplied by  $p+2$ after each iteration (when the procedure converges, of course). For $p=0$ and $p=1$, one obtains
Newton's and Halley's schemes, respectively. The scheme (\ref{solwexp}) for solving the Colebrook equation is obtained with $p=2$ together with the function $f$ given by the equation (\ref{eqom}), plus some elementary algebra. Intensive tests have convinced us that it is most probably the best choice for the problem at hand here.

\section{MATLAB code}\label{appmat}

The {\sc Matlab${}^\copyright$} function below is a vectorized implementation of the algorithm described in this paper. This code can also be freely downloaded \cite{cla}. We hope that the program is sufficiently documented so that one can easily test and modify it.

\lstset{basicstyle=\scriptsize}
\begin{lstlisting}

function F = colebrook(R,K)
% F = COLEBROOK(R,K) fast, accurate and robust computation of the
% Darcy-Weisbach friction factor according to the Colebrook formula:
%                            -                       -
%      1                    |    K        2.51        |
%  ---------  =  -2 * Log10 |  ----- + -------------  |
%   sqrt(F)                 |   3.7     R * sqrt(F)   |
%                            -                       -
% INPUT:
%   R : Reynolds' number (should be > 2300).
%   K : Equivalent sand roughness height divided by the hydraulic
%       diameter (default K=0).
%
% OUTPUT:
%   F : Friction factor.
%
% FORMAT:
%   R, K and F are either scalars or compatible arrays.
%
% ACCURACY:
%   Around machine precision for all R > 3 and for all 0 <= K,
%   i.e. in an interval exceeding all values of physical interest.
%
% EXAMPLE: F = colebrook([3e3,7e5,1e100],0.01)

% Check for errors.
if any(R(:)<=0) == 1,
   error('The Reynolds number must be positive (R>2000).');
end,
if nargin == 1,
   K = 0;
end,
if any(K(:)<0) == 1,
   error('The relative sand roughness must be non-negative.');
end,

% Initialization.
X1 = K .* R * 0.123968186335417556;     % X1 <- K * R * log(10) / 18.574.
X2 = log(R) - 0.779397488455682028;     % X2 <- log( R * log(10) / 5.02 );

% Initial guess.
F = X2 - 0.2;                           % F <- X2 - 1/5;

% First iteration.
E = ( log(X1+F) + F - X2 ) ./ ( 1 + X1 + F );
F = F - (1+X1+F+0.5*E) .* E .* (X1+F) ./ (1+X1+F+E.*(1+E/3));

% Second iteration (remove the next two lines for moderate accuracy).
E = ( log(X1+F) + F - X2 ) ./ ( 1 + X1 + F );
F = F - (1+X1+F+0.5*E) .* E .* (X1+F) ./ (1+X1+F+E.*(1+E/3));

% Finalized solution.
F = 1.151292546497022842 ./ F;          % F <- 0.5 * log(10) / F;
F = F .* F;                             % F <- Friction factor.

\end{lstlisting}

\section{FORTRAN code}\label{appfor}

The FORTRAN function below was written with maximum speed in mind, so some trivial arithmetic simplifications were used and there are no check for errors in the input parameters.

\lstset{basicstyle=\scriptsize}
\begin{lstlisting}

      DOUBLE PRECISION FUNCTION COLEBROOK(R,K)

C  F = COLEBROOK(R,K) computes the Darcy-Weisbach friction
C  factor according to the Colebrook-White formula.
C
C  R : Reynold's number.
C  K : Roughness height divided by the hydraulic diameter.
C  F : Friction factor.

      IMPLICIT NONE
      DOUBLE PRECISION R, K, F, E, X1, X2, T
      PARAMETER ( T = 0.333333333333333333D0 )

C Initialization.
      X1 = K * R * 0.123968186335417556D0
      X2 = LOG(R) - 0.779397488455682028D0

C Initial guess.
      F = X2 - 0.2D0

C First iteration.
      E = (LOG(X1+F)-0.2D0) / (1.0D0+X1+F)
      F = F - (1.0D0+X1+F+0.5D0*E)*E*(X1+F) / (1.0D0+X1+F+E*(1.0D0+E*T))

C Second iteration (if needed).
      IF ((X1+X2).LT.(5.7D3)) THEN
         E = (LOG(X1+F)+F-X2) / (1.0D0+X1+F)
         F = F - (1.0D0+X1+F+0.5D0*E)*E*(X1+F) / (1.0D0+X1+F+E*(1.0D0+E*T))
      ENDIF

C Finalized solution.
      F = 1.151292546497022842D0 / F
      COLEBROOK = F * F

      RETURN
      END

\end{lstlisting}

Note that, depending on the FORTRAN version and on the compiler, the command {\sf\small LOG} may have to be replaced by {\sf\small DLOG} to ensure that the logarithm is computed with a double-precision accuracy. \\

\newpage


\newpage

\newpage

\end{document}